\documentclass[pre,twocolumn,showpacs,eqsecnum]{revtex4}
\begin{document}
\title{Continuous-time random walk with a superheavy-tailed
distribution of waiting times}
\author{S.~I.~Denisov$^{1}$}
\email{denisov@sumdu.edu.ua}
\author{H.~Kantz$^{2}$}
\affiliation{$^{1}$Sumy State University, Rimsky-Korsakov Street 2, UA-40007
Sumy, Ukraine\\
$^{2}$Max-Planck-Institut f\"{u}r Physik komplexer Systeme, N\"{o}thnitzer
Stra{\ss}e 38, D-01187 Dresden, Germany}


\begin{abstract}
We study the long-time behavior of the probability density associated with the
decoupled continuous-time random walk which is characterized by a
superheavy-tailed distribution of waiting times. It is shown that if the random
walk is unbiased (biased) and the jump distribution has a finite second moment
then the properly scaled probability density converges in the long-time limit
to a symmetric two-sided (an asymmetric one-sided) exponential density. The
convergence occurs in such a way that all the moments of the probability
density grow slower than any power of time. As a consequence, the reference
random walk can be viewed as a generic model of superslow diffusion. A few
examples of superheavy-tailed distributions of waiting times that give rise to
qualitatively different laws of superslow diffusion are considered.
\end{abstract}
\pacs{05.40.Fb, 02.50.Ey, 02.50.Fz}

\maketitle

\section{INTRODUCTION}
\label{Intr}

Almost a half century ago, Montroll and Weiss \cite{MW} introduced a special
class of cumulative processes, the so-called continuous-time random walks
(CTRWs). Due to its simplicity and flexibility, the CTRW approach has become an
important tool for the analysis of various stochastic systems. It is especially
efficient for studying the systems whose average behavior differs from the
classical one characterized by the normal, linear dependence of the mean-square
displacement (variance) on time. In particular, the CTRW approach is
extensively used to determine the laws of anomalous diffusion occurring in
these systems (see, e.g., Refs.~\cite{BG,MK,AH}).

One of the most important characteristics of the CTRW is the long-time behavior
of the probability density $P(x,t)$ associated with the position $x(t)$ of a
walking particle. It is this density that describes in full detail the
diffusion properties of particles, including the asymptotic behavior of the
variance (i.e., the diffusion law). Moreover, the probability density $P(x,t)$
at $t\to \infty$ plays a significant role in establishing the connection
between the CTRW and the fractional time diffusion equation \cite{BS}.

Within the CTRW framework, the probability density $P(x,t)$ of the particle
position is completely characterized by the joint probability density of the
waiting times $\tau_{n}$ ($n=1,2,\ldots$), i.e., random times between the
particle jumps, and the jump magnitudes $x_{n}$. For the decoupled CTRW, when
the sets $\{ \tau_{n}\}$ and $\{ x_{n}\}$ are statistically independent of each
other, $P(x,t)$ depends only on the probability densities $p(\tau)$ and $w(x)$
of $\tau_{n}$ and $x_{n}$, respectively. In this case the long-time behavior of
$P(x,t)$ is determined by the main characteristics of $p(\tau)$ and $w(x)$
\cite{Tun,SKW,Kot}. For example, if the second moment of $w(x)$ exists and
$p(\tau)$ at $\tau \to \infty$ is given by the asymptotic formula $p(\tau) \sim
h/\tau^{1 + \alpha}$ with $h$ being a positive parameter and $0<\alpha<1$, then
the properly scaled probability density $P(x,t)$ at $t\to \infty$ converges
either to a symmetric two-sided (for an unbiased walk) or to an asymmetric
one-sided (for a biased walk) stable density \cite{Tun}.

A more general class of waiting-time probability densities is described by the
asymptotic relation $p(\tau) \sim h(\tau)/\tau^{1 + \alpha}$ ($\tau \to
\infty$), where a positive function $h(\tau)$ is assumed to be slowly varying
at infinity. Due to this property the function $h(\tau)$ does not strongly
influence the long-time behavior of $x(t)$ if $\alpha>0$ \cite{Shles}. However,
at $\alpha=0$ the situation changes drastically. From a mathematical point of
view, this is a consequence of the fact that the waiting-time densities
characterized by the condition $\alpha=0$ belong to a class of
superheavy-tailed densities \cite{DKH,DK1} whose all fractional moments are
infinite. It is expected, therefore, that in this case the long-time behavior
of the CTRW is mainly controlled by the asymptotic form of slowly varying
function $h(\tau)$ at $\tau \to \infty$. The laws of both biased and unbiased
superslow diffusion \cite{DK2}, i.e., diffusion of particles whose mean-square
displacement grows slower than any power of time, corroborate this statement.

Some important features of the long-time behavior of the probability density
$P(x,t)$ have already been established by Havlin and Weiss \cite{HW1} for a
very special class of slowly varying functions with $h(\tau) \sim a/\ln^{1 +
\nu} \tau$ ($a>0$, $\nu>0$) as $\tau \to \infty$. They have shown, in
particular, that if the CTRW is decoupled and unbiased and the second moment of
the jump probability density $w(x)$ is finite then $P(x,t)$ at $t \to \infty$
has symmetric exponential tails. In contrast, in the present paper we provide a
detailed analysis of the long-time behavior of $P(x,t)$ for a whole class of
slowly varying functions $h(\tau)$ that support the normalization condition of
$p(\tau)$ and consider both biased and unbiased versions of the decoupled CTRW.

The paper is organized as follows. In Sec.~\ref{Spec}, we reproduce the basic
equations of the decoupled CTRW and specify a class of superheavy-tailed
densities of waiting times. The main results of the paper are obtained in
Sec.~\ref{LTB}. Here, we determine the long-time behavior of the probability
density of the particle position and establish its limiting form for a properly
scaled spatial variable. Moreover, we use the two-sided exponential jump
density to verify our results and consider a few illustrative examples of
superheavy-tailed densities of waiting times. In Sec.~\ref{Mom}, we calculate
the moments of the particle position at long times and show that they grow
slower than any power of time. In the same section, we derive the most general
form of the laws of superslow diffusion and calculate the laws that correspond
to the illustrative examples. Our findings are summarized in Sec.~\ref{Concl}.

\section{SPECIFICATION OF THE MODEL}
\label{Spec}

The CTRW is defined as the cumulative continuous-time jump process
\cite{BG,MK,AH}
\begin{equation}
    X(t) = \sum_{n=1}^{N(t)} x_{n},
\label{X(t)}
\end{equation}
where $X(t)$ ($X(0)=0$) can be considered as the walker (particle) position,
$x_{n} \in (-\infty, \infty)$ is the random magnitude of the $n$th jump of a
particle, and $N(t)= 0,1,2,\ldots$ is the random number of jumps up to time $t$
(if $N(t)=0$ then $X(t)=0$). It is assumed that both the magnitudes $x_{n}$ and
the waiting times $\tau_{n}$ (i.e., times between successive jumps) are
independent random variables distributed with the probability densities $w(x)$
and $p(\tau)$, respectively. In the decoupled case, when the sets $\{ x_{n} \}$
and $\{ \tau_{n} \}$ of the variables $x_{n}$ and $\tau_{n}$ are independent of
each other, the distribution of particles is completely determined by these
densities. Specifically, in the Fourier-Laplace space the probability density
$P(x,t)$ of the particle position $X(t)$ is given by the Montroll-Weiss
equation \cite{MW}
\begin{equation}
    P_{ks} = \frac{1-p_{s}}{s(1-p_{s}w_{k})}.
\label{M-Weq}
\end{equation}
Here, the indexes $k$ and $s$ denote the Fourier and Laplace transforms, $u_{k}
= \mathcal{F} \{u(x)\} = \int_{-\infty}^{\infty} dx e^{ikx} u(x)$ and $v_{s} =
\mathcal{L} \{v(t)\} = \int_{0}^{\infty} dt e^{-st} v(t)\; (\mathrm{Re} s>0)$,
of the corresponding functions. Applying to Eq.~(\ref{M-Weq}) the inverse
Fourier and Laplace transforms defined as $u(x) = \mathcal{F}^{-1} \{u_{k}\} =
(1/2\pi) \int_{-\infty}^{\infty} dk e^{-ikx} u_{k}$ and $v(t) =
\mathcal{L}^{-1} \{v_{s}\} = (1/2\pi i) \int_{c- i\infty}^{c+ i\infty} ds
e^{st} v_{s}$ (the real parameter $c$ is assumed to be larger than the real
parts of all singularities of $v_{s}$), respectively, one obtains
\begin{equation}
    P(x,t) = \mathcal{L}^{-1} \bigg\{ \frac{1-p_{s}}{s}\,
    \mathcal{F}^{-1} \bigg\{ \frac{1}{1-p_{s}w_{k}} \bigg\} \bigg\}.
\label{P}
\end{equation}

It can also be shown \cite{KMS,MBK,MRGS} that the probability density $P(x,t)$
satisfies the integral equation
\begin{eqnarray}
    P(x,t) \!&=&\! V(t)\delta(x) + \int_{0}^{t}d\tau
    \int_{-\infty}^{\infty} dx'P(x',\tau)
    \nonumber\\[6pt]
    &&\!\! \times\, p(t-\tau) w(x-x'),
    \label{Peq}
\end{eqnarray}
where $\delta (x)$ is the Dirac $\delta$ function and
\begin{equation}
    V(t)= 1 -\int_{0}^{t}d\tau p(\tau) = \int_{t}^{\infty}d\tau p(\tau)
\label{defV}
\end{equation}
($V(0)=1$, $V(\infty)=0$) is the complementary cumulative distribution function
of waiting times, that is also known as the survival or exceedance probability.
The first term in the right-hand side of this equation corresponds to the
situation, which is realized with probability $V(t)$, when there are no jumps
up to time $t$. While the exact solution of Eq.~(\ref{Peq}) was obtained for a
special case of jump and waiting-time densities \cite{Bar}, the long-time
behavior of $P(x,t)$ was analyzed for a much wider class of these densities.
This has been done by using the representation (\ref{P}) \cite{Tun,SKW} and by
applying the limit theorems of probability theory \cite{Kot,MS}. The first
approach is based on the Tauberian theorem for the Laplace transforms (see,
e.g., Ref.~\cite{Fel}), which states that if the function $v(t)$ is ultimately
monotone and
\begin{equation}
    v_{s} \sim \frac{1}{s^{\gamma}}\, L\!\left( \frac{1}{s} \right)
\label{vs1}
\end{equation}
($0<\gamma<\infty$) as $s \to 0$ then
\begin{equation}
    v(t) \sim \frac{t^{\gamma-1}}{\Gamma(\gamma)}\, L (t)
\label{vt1}
\end{equation}
as $t \to \infty$. Here, $\Gamma(\gamma)$ is the gamma function and $L(t)$ is a
slowly varying function at infinity, i.e., $L(\mu t) \sim L(t)$ as $t \to
\infty$ for all $\mu>0$. We note that, in contrast to the inverse Laplace
transform where the parameter $s$ is complex with $\mathrm{Re} s>0$, in
Eq.~(\ref{vs1}) this parameter is assumed to be real and positive.

In this paper, we use the Laplace transform
\begin{equation}
    P_{s}(x) = \frac{1-p_{s}}{s}\, \delta(x) +
    \frac{(1-p_{s})p_{s}}{s}\, \mathcal{F}^{-1} \bigg\{
    \frac{w_{k}}{1-p_{s}w_{k}} \bigg\},
\label{P(x)s}
\end{equation}
which follows from Eq.~(\ref{P}), and the Tauberian theorem to study the
long-time behavior of the probability density $P(x,t)$ in the case of
superheavy-tailed distributions of waiting times. More precisely, we consider a
class of probability distributions whose densities have the following
asymptotic behavior:
\begin{equation}
    p(\tau) \sim \frac{h(\tau)}{\tau}
\label{p as}
\end{equation}
($\tau \to \infty$), where a positive function $h(\tau)$ varies slowly at
infinity. Since for each slowly varying function $h(\tau)$ the condition
$\tau^{\rho}h(\tau) \to \infty$ ($\rho>0$) holds as $\tau \to \infty$
\cite{BGT}, all fractional moments of $p(\tau)$ are infinite [i.e.,
$\mathcal{T}_{\rho} = \int_{0}^{\infty} d\tau \tau^{\rho} p(\tau) = \infty$].
It should be noted that the slowly varying function $h(\tau)$ in Eq.~(\ref{p
as}) is not arbitrary: it must be compatible with the normalization condition
$\mathcal{T}_{0} = \int_{0}^{\infty}d\tau p(\tau) = 1$, which in turn implies
that $h(\tau) = o(1/\ln\tau)$ as $\tau \to \infty$. Finally, we assume here
that the first two moments of the jump density $w(x)$, $l_{1} = \int_{-\infty}^
{\infty} dxxw(x)$ and $l_{2} = \int_{-\infty}^ {\infty} dxx^{2}w(x)$, exist. As
will be shown below, in this case only these characteristics of $w(x)$
influence the long-time behavior of $P(x,t)$.

\section{LONG-TIME BEHAVIOR OF $\textit{P}\textbf{(}\textit{x,t}
\textbf{)}$}
\label{LTB}

According to the Tauberian theorem, the long-time behavior of the probability
density $P(x,t)$ of the particle position is determined by the asymptotic
behavior of the Laplace transform $P_{s}(x)$, see Eq.~(\ref{P(x)s}), at $s \to
0$. In order to find $P_{s}(x)$ for small $s$, let us first determine the two
terms of the asymptotic expansion of the Laplace transform $p_{s} =
\int_{0}^{\infty} d\tau e^{-s\tau}p(\tau)$. With this purpose, we use the
exceedance probability (\ref{defV}) to represent $p_{s}$ as
\begin{equation}
    p_{s} = 1 - \int_{0}^{\infty}dq e^{-q} V\! \left( \frac{q}{s}
    \right)\!.
\label{ps2}
\end{equation}
An important feature of the exceedance probability, which follows from the
asymptotic formula (\ref{p as}), is that it is a slowly varying function at
infinity, i.e., $V(\mu t) \sim V(t)$ as $t\to \infty$ \cite{DK2}. Therefore, at
$s\to 0$ the function $V(q/s)$ in Eq.~(\ref{ps2}) can be replaced by $V(1/s)$,
yielding
\begin{equation}
    p_{s} \sim 1 - V\! \left( \frac{1}{s} \right)\!.
\label{ps3}
\end{equation}

Since $p_{s}$ and $w_{k}$ tend to 1 as $s$ and $k$ tend to 0, at small $s$ the
main contribution to the inverse Fourier transform in Eq.~(\ref{P(x)s}) comes
from the close vicinity of the point $k=0$. Due to this fact, $w_{k}$ can be
approximated by a few terms of the expansion of $w_{k}$ as $k \to 0$. Taking
into account that the first two moments of $w(x)$, $l_{1}$ and $l_{2}$, are
assumed to exist, we obtain
\begin{equation}
    w_{k} \sim 1 + il_{1}k - \frac{1}{2}\, l_{2}k^{2}
\label{wk}
\end{equation}
($k \to 0$). Then, using Eqs.~(\ref{ps3}) and (\ref{wk}), we can find the
asymptotic behavior of $\mathcal{F}^{-1}\{ w_{k}/(1- p_{s}w_{k})\}$ as $s\to 0$
by replacing $w_{k}/(1- p_{s}w_{k})$ by $[l_{2}k^{2}/2 + V(1/s)]^{-1}$ in the
unbiased case (when $l_{1}=0$) and by $[- il_{1}k + V(1/s)]^{-1}$ in the biased
case (when $l_{1} \neq 0$).

For the former case one has \cite{Erd}
\begin{eqnarray}
    \mathcal{F}^{-1} \bigg\{ \frac{w_{k}}{1-p_{s}w_{k}} \bigg\}
    \!&\sim&\! \mathcal{F}^{-1} \bigg\{ \frac{1}{l_{2}k^{2}/2 +
    V(1/s)} \bigg\}
    \nonumber\\[6pt]
    \!&=&\! \frac{e^{-|x|\sqrt{2V(1/s)/l_{2}}}}{\sqrt{2l_{2}V(1/s)}}.
\label{F1}
\end{eqnarray}
According to this result, the first term in the right-hand side of
Eq.~(\ref{P(x)s}) can be neglected in the limit $s \to 0$. Indeed, the integral
of this term over some interval containing the point $x=0$ is equal to $(1
-p_{s})/s \sim V(1/s)/s$, while Eq.~(\ref{F1}) shows that the integral of the
second term is proportional to $1/s$. Therefore, since $V(1/s) \to 0$ as $s \to
0$, the main term of the asymptotic expansion of $P_{s}(x)$ takes the form
\begin{equation}
    P_{s}(x) \sim \frac{1}{s} \, \sqrt{\frac{V(1/s)}{2l_{2}}}\,
    e^{-|x|\sqrt{2V(1/s)/l_{2}}}
\label{Ps1}
\end{equation}
($s \to 0$). Because the exceedance probability $V(t)$ varies slowly at
infinity, it is not difficult to show that so also does the function
$\sqrt{V(t) /2l_{2}}\, e^{-|x|\sqrt{2V(t) /l_{2}}}$. Hence, the Tauberian
theorem is applicable to this case and, in accordance with Eq.~(\ref{Ps1}),
leads to the following behavior of $P(x,t)$ at long times:
\begin{equation}
    P(x,t) \sim \sqrt{\frac{V(t)}{2l_{2}}}\,
    e^{-|x|\sqrt{2V(t)/l_{2}}},
\label{P1}
\end{equation}
where, in accordance with Eq.~(\ref{p as}),
\begin{equation}
    V(t) \sim \int_{t}^{\infty}d\tau \frac{1}{\tau} h(\tau).
\label{V1}
\end{equation}

It should be stressed that in spite of the fact that the asymptotic formula
(\ref{P1}) is derived considering a small vicinity of the point $k=0$, it is
valid for all $x$. The reason is that it is this vicinity which is responsible
for the asymptotic behavior of $P_{s}(x)$ as $s \to 0$. We note in this context
that the asymptotic formula for $P(x,t)$ obtained in \cite{HW1} for a special
case of the exceedance probability is applicable for all $x$, not only for $|x|
\to \infty$ as it was assumed in Ref.~\cite{HW1}. Finally, introducing the
variable $y=x\sqrt{2V(t) /l_{2}}$, we make sure that the limiting distribution
of the scaled particle position $Y(t)=X(t) \sqrt{2V(t)/l_{2}}$ is described by
the symmetric two-sided exponential density
\begin{equation}
    \mathcal{P}(y) = \lim_{t \to \infty}  \sqrt{\frac{l_{2}}
    {2V(t)}}\,P\!\left( \sqrt{\frac{l_{2}}{2V(t)}}\,y,t \right)
    = \frac{e^{-|y|}}{2}.
\label{limP1}
\end{equation}

For the biased CTRW, when $l_{1} \neq 0$, we obtain \cite{Erd}
\begin{eqnarray}
    \mathcal{F}^{-1} \bigg\{ \frac{w_{k}}{1-p_{s}w_{k}} \bigg\}
    \!&\sim&\! \mathcal{F}^{-1} \bigg\{ \frac{1}{- il_{1}k +
    V(1/s)} \bigg\}
    \nonumber\\[6pt]
    \!&=&\! \frac{1}{|l_{1}|}\, e^{-xV(1/s)/l_{1}} \theta(l_{1}x),
\label{F2}
\end{eqnarray}
where $\theta(x) = 0$ if $x<0$ and $\theta(x)=1$ if $x>0$. It is important to
emphasize that the condition $\mathcal{F}^{-1} \{ w_{k}/(1-p_{s}w_{k}) \}
\approx 0$, which occurs at $l_{1}x<0$, is a direct consequence of the used
approximation [see the first line of Eq.~(\ref{F2})]. In fact, the term
$[-il_{1}k + V(1/s)]^{-1}$ is the principal part of the Laurent expansion of $
w_{k}/(1- p_{s}w_{k})$ in the vicinity of the point $k=-iV(1/s)/l_{1} \to 0$.
Taking the next term of this expansion, $[il_{1}k + c]^{-1}$ ($c>0$), which is
responsible for the behavior of $\mathcal{F}^{-1} \{ w_{k}/(1-p_{s}w_{k}) \}$
at $l_{1}x<0$, one gets $\mathcal{F}^{-1} \{ w_{k}/(1-p_{s}w_{k}) \} \sim
e^{xc/l_{1}}/|l_{1}|$ ($xl_{1}<0$). However, since for small $s$ the condition
$c/V(1/s) \gg 1$ holds, $\mathcal{F}^{-1} \{ w_{k}/(1-p_{s}w_{k}) \}$ as a
function of $x$ at $l_{1}x>0$ varies much slower than it does for $l_{1}x<0$.
Therefore, in the limit $s \to 0$ the contribution coming from the region with
$l_{1}x<0$ is negligible, and the inverse Fourier transform in
Eq.~(\ref{P(x)s}) can be safely approximated by Eq.~(\ref{F2}) (see also
Sec.~\ref{Test}).

For the same reason as before, in the biased case the first term in the
right-hand side of Eq.~(\ref{P(x)s}) can also be neglected in the limit $s \to
0$. Therefore, using Eqs.~(\ref{P(x)s}), (\ref{ps3}) and (\ref{F2}), one gets
\begin{equation}
    P_{s}(x) \sim  \frac{1}{s|l_{1}|} V\! \left( \frac{1}{s}
    \right)\! e^{-xV(1/s)/l_{1}} \theta(l_{1}x)
\label{Ps2}
\end{equation}
($s \to 0$), and the Tauberian theorem yields
\begin{equation}
    P(x,t) \sim \frac{V(t)}{|l_{1}|} \,e^{-xV(t)/l_{1}}
    \theta(l_{1}x)
\label{P2}
\end{equation}
($t \to \infty$). Accordingly, the limiting distribution of the scaled particle
position $Y(t) = X(t)V(t)/l_{1}$ is described by the asymmetric one-sided
exponential density
\begin{equation}
    \mathcal{P}(y) = \lim_{t \to \infty} \frac{|l_{1}|}
    {V(t)}\,P\!\left( \frac{l_{1}}{V(t)}\,y,t \right)
    = e^{-y} \theta(y).
\label{limP2}
\end{equation}

Equations (\ref{limP1}) and (\ref{limP2}) are the main results of this paper.
They show that in the case of the decoupled CTRWs with superheavy-tailed
distributions of waiting times the limiting distribution of the properly scaled
particle position is described by either the two-sided or the one-sided
exponential density. Remarkably, these two- and one-sided exponential densities
correspond to all unbiased and all biased CTRWs, respectively.

\subsection{Testing jump density}
\label{Test}

The limiting probability densities (\ref{limP1}) and (\ref{limP2}) were
obtained from the approximation of the inverse Fourier transform in
Eq.~(\ref{P(x)s}) by the asymptotic formulas (\ref{F1}) and (\ref{F2}).
Although the applicability of this approximation is well-grounded, it is
desirable to check the validity of Eqs.~(\ref{limP1}) and (\ref{limP2}) by
using an exact expression for that transform. To this end, we consider the
two-sided exponential jump density
\begin{equation}
    w(x) = \frac{\kappa_{+}\kappa_{-}}{\kappa_{+} + \kappa_{-}}
    \left\{\! \begin{array}{ll}
    e^{-x\kappa_{+}}, & x \geq 0
    \\ [6pt]
    e^{x\kappa_{-}}, & x < 0
    \end{array}
    \right.
\label{w}
\end{equation}
($\kappa_{\pm}\!>0$). Taking into account that the Fourier transform of this
density is given by
\begin{equation}
    w_{k} = \frac{\kappa_{+}\kappa_{-}}{(k + i\kappa_{+})
    (k - i\kappa_{-})},
\label{wk2}
\end{equation}
we obtain
\begin{equation}
    \frac{w_{k}}{1-p_{s}w_{k}} = \frac{\kappa_{+}\kappa_{-}}
    {(k + i\tilde{\kappa}_{+})(k - i\tilde{\kappa}_{-})},
\label{rel1}
\end{equation}
where
\begin{equation}
    \tilde{\kappa}_{\pm} = \mp\frac{1}{2}(\kappa_{-} - \kappa_{+})
    + \frac{1}{2} \sqrt{(\kappa_{-} - \kappa_{+})^{2} +
    4(1-p_{s})\kappa_{+}\kappa_{-}}.
\label{kap}
\end{equation}
Since the right-hand sides of Eqs.~(\ref{wk2}) and (\ref{rel1}) are similar,
the inverse Fourier transform of $w_{k}/ (1-p_{s}w_{k})$ is of the form of
$w(x)$; that is,
\begin{equation}
    \mathcal{F}^{-1} \bigg\{ \frac{w_{k}}{1-p_{s}w_{k}} \bigg\} =
    \frac{\kappa_{+}\kappa_{-}}
    {\tilde{\kappa}_{+} + \tilde{\kappa}_{-}}
    \left\{\! \begin{array}{ll}
    e^{-x\tilde{\kappa}_{+}}, & x \geq 0
    \\ [6pt]
    e^{x\tilde{\kappa}_{-}}, & x < 0,
    \end{array}
    \right.
\label{F3}
\end{equation}
and so Eq.~(\ref{P(x)s}) in the reference case reads
\begin{eqnarray}
    P_{s}(x) \!&=&\! \frac{1-p_{s}}{s}\, \delta(x) +
    \frac{(1-p_{s})p_{s}}{s}\, \frac{\kappa_{+}\kappa_{-}}
    {\tilde{\kappa}_{+} + \tilde{\kappa}_{-}}
    \nonumber\\[6pt]
    &&\! \times \left\{\!
    \begin{array}{ll}
    e^{-x\tilde{\kappa}_{+}}, & x \geq 0
    \\ [6pt]
    e^{x\tilde{\kappa}_{-}}, & x < 0.
    \end{array}
    \right.
\label{Ps3}
\end{eqnarray}

If $\kappa_{+} = \kappa_{-} = \kappa$ then Eq.~(\ref{w}) yields $l_{1} =0$
(i.e., the corresponding CTRW is unbiased). In this case $\tilde{\kappa}_{+} =
\tilde{\kappa}_{-} = \kappa \sqrt{1-p_{s}} \sim \kappa V^{1/2}(1/s)$, $l_{2} =
2/\kappa^{2}$, and Eq.~(\ref{Ps3}) at $s \to 0$ reduces to Eq.~(\ref{Ps1}).
Thus, the limiting probability density which corresponds to Eq.~(\ref{Ps3})
with $\kappa_{+} = \kappa_{-}$ is given by Eq.~(\ref{limP1}), as it should be
for the unbiased case. If $\kappa_{+} \neq \kappa_{-}$ then the biased CTRW
with $l_{1} = (\kappa_{-} - \kappa_{+})/\kappa_{+}\kappa_{-}$ occurs. According
to Eq.~(\ref{kap}), in this case we have
\begin{equation}
    \tilde{\kappa}_{\pm} \sim \kappa_{+}\kappa_{-}|l_{1}|
    \theta(\mp l_{1}) + \frac{1}{|l_{1}|}V\! \left(\frac{1}{s}\right)\!
    \theta(\pm l_{1})
\label{rel2}
\end{equation}
as $s \to 0$. Using this result and Eq.~(\ref{Ps3}), it is not difficult to
find the asymptotic formula for $P_{s}(x)$ which, by applying the Tauberian
theorem, gives
\begin{equation}
    P(x,t) \sim \frac{\kappa_{+}\kappa_{-}|l_{1}|V(t)}
    {\kappa_{+}\kappa_{-}|l_{1}|^{2} + V(t)} \!
    \left\{\! \begin{array}{ll}
    e^{-xV(t)/l_{1}}, & l_{1}x > 0
    \\ [6pt]
    e^{x\kappa_{+}\kappa_{-}l_{1}}, & l_{1}x < 0
    \end{array}
    \right.
\label{P3}
\end{equation}
as $t \to \infty$. While at $l_{1}x>0$ the asymptotic formulas (\ref{P3}) and
(\ref{P2}) are asymptotically equivalent, at $l_{1}x<0$ they are different.
However, as it was mentioned earlier, this difference does not affect the
long-time distribution of the scaled particle position $Y(t) = X(t)V(t)/l_{1}$.
Indeed, taking into account that the exponential term $e^{y\kappa_{+}
\kappa_{-}l^{2}_{1} /V(t)}$ at $y<0$ tends to zero as $t\to \infty$, the
limiting probability density $\mathcal{P}(y)$ associated with probability
density $P(x,t)$ from Eq.~(\ref{P3}) is given by Eq.~(\ref{limP2}).

\subsection{Illustrative examples of waiting-time densities}
\label{Ex}

The long-time behavior of the exceedance probability $V(t)$ is the most
important characteristic of the CTRWs with superheavy-tailed distributions of
waiting times. As follows from Eqs.~(\ref{P1}) and (\ref{P2}), it is this
behavior that is responsible for the long-time behavior of the probability
density $P(x,t)$. Below we consider three illustrative examples of the
waiting-time probability density $p(\tau)$ which lead to different asymptotic
formulas for $V(t)$.

As a first example, we consider the probability density
\begin{equation}
    p(\tau) = \frac{qb^{1/q}}{\Gamma(1/q, b\ln^{q} \eta)}
    \frac{\exp[-b\ln^{q}(\eta + \tau)]}{\eta + \tau},
\label{p1}
\end{equation}
where $\Gamma(a,x) = \int_{x}^{\infty} dy e^{-y} y^{a-1}$ is the upper
incomplete gamma function [$\Gamma(a,0) =\Gamma(a)$], and the conditions
$\eta>1$, $b>0$ and $0<q<0$ are assumed to hold. The conditions $\eta>1$, $b>0$
and $q>0$ are responsible for the positivity and normalization of $p(\tau)$,
and the condition $q<1$ guarantees that $h(\tau)$ is a slowly varying function.
The last inequality follows from that the limit
\begin{equation}
    \lim_{\tau \to \infty} \frac{h(\mu\tau)}{h(\tau)} =
    \lim_{\tau \to \infty} \exp(-bq \ln\mu \ln^{q-1}\tau)
\label{lim1}
\end{equation}
equals 1 for all $\mu>0$, i.e., the function $h(\tau)$ is slowly varying at
infinity, only if $q<1$. According to the definition (\ref{defV}), the
exceedance probability that corresponds to the probability density (\ref{p1})
reads
\begin{equation}
    V(t) = \frac{\Gamma[1/q, b\ln^{q}(\eta + t)]}
    {\Gamma(1/q, b\ln^{q} \eta)}.
\label{V3}
\end{equation}
Therefore, using the asymptotic formula \cite{AS} $\Gamma(a,x) \sim e^{-x}
x^{a-1}$ as $x\to \infty$, one gets in the long-time limit
\begin{equation}
    V(t) \sim \frac{b^{1/q -1}} {\Gamma(1/q, b\ln^{q} \eta)}
    \exp(-b\ln^{q} t)\ln^{1-q} t.
\label{V3as}
\end{equation}

Our second example is the probability density
\begin{equation}
    p(\tau) = \frac{(r-1)\ln^{r-1}\eta}{(\eta + \tau)
    \ln^{r} (\eta + \tau)},
\label{p2}
\end{equation}
where $r>1$ and $\eta>1$. The main feature of this density is that its right
tail is heavier than in the previous case; that is, $p(\tau)$ at $\tau \to
\infty$ tends to zero slower than that in Eq.~(\ref{p1}). The exceedance
probability associated with the probability density (\ref{p2}) has the form
\begin{equation}
    V(t) = \bigg(\frac{\ln \eta}{\ln(\eta+t)}\bigg)^{r-1},
\label{V4}
\end{equation}
and so
\begin{equation}
    V(t) \sim \bigg(\frac{\ln \eta}{\ln t}\bigg)^{r-1}
\label{V4as}
\end{equation}
as $t \to \infty$. Comparing the asymptotic formulas (\ref{V3as}) and
(\ref{V4as}), we conclude that the exceedance probability in Eq.~(\ref{V4})
decreases slower than the exceedance probability in Eq.~(\ref{V3}).

Finally, as a third example, we consider the waiting-time probability density
\begin{equation}
    p(\tau) = \frac{(r-1)(\ln\ln \eta)^{r-1}}{(\eta + \tau)
    \ln(\eta + \tau) [\ln\ln (\eta + \tau)]^{r}}
\label{p3}
\end{equation}
($\eta>e$, $r>1$), whose right tail is heavier than for $p(\tau)$ from
Eq.~(\ref{p2}). In this case the exceedance probability is also calculated
exactly, yielding
\begin{equation}
    V(t) = \bigg(\frac{\ln\ln \eta}{\ln\ln(\eta+t)}\bigg)^{r-1}
\label{V5}
\end{equation}
and
\begin{equation}
    V(t) \sim \bigg(\frac{\ln\ln \eta}{\ln\ln t}\bigg)^{r-1}
\label{V5as}
\end{equation}
as $t \to \infty$. We note that the above examples are illustrative and they do
not exhaust all possible behaviors of the exceedance probability $V(t)$ at long
times. Moreover, a class of slowly varying functions $V(t)$ contains infinitely
many functions that, at $t \to \infty$, decrease both faster and slower than a
given function from this class.

\section{MOMENTS OF THE PARTICLE POSITION}
\label{Mom}

The moments $M_{n}(t)$ ($n=1,2,\ldots$) of the particle position are defined in
the usual way:
\begin{equation}
    M_{n}(t) = \int_{-\infty}^{\infty} dx x^{n}P(x,t).
\label{defMn}
\end{equation}
Their behavior at long times can be easily determined by using the limiting
probability densities (\ref{limP1}) and (\ref{limP2}). Specifically, in the
case of unbiased CTRWs for the even moments we obtain
\begin{eqnarray}
    M_{2n}(t) \!&=&\! \left(\frac{l_{2}}{2V(t)}\right)^{n+1/2}
    \!\int_{-\infty}^{\infty}dy y^{2n}P\!\left( \sqrt{\frac{l_{2}}
    {2V(t)}}\,y,t \right)
    \nonumber\\[6pt]
    \!&\sim&\! \left(\frac{l_{2}}{2V(t)}\right)^{n}\!
    \int_{0}^{\infty}dy y^{2n}e^{-y}.
\label{M2n1}
\end{eqnarray}
Taking into account that $\int_{0}^{\infty}dy y^{2n}e^{-y} = \Gamma(2n+1) =
(2n)!$, the above result becomes
\begin{equation}
    M_{2n}(t) \sim \! \left( \frac{l_{2}}{2V(t)} \right)^{\!n}(2n)!.
\label{M2n2}
\end{equation}
Since in this case $P(x,t)$ is a symmetric function of $x$, all odd moments
equal zero: $M_{2n-1}(t)=0$.

For the biased CTRWs Eqs.~(\ref{defMn}) and (\ref{limP2}) give
\begin{eqnarray}
    M_{n}(t) \!&=&\! \left( \frac{l_{1}}{V(t)} \right)^{n+1}
    \!\int_{-l_{1}\infty}^{l_{1}\infty}dy y^{n}P\!
    \left( \frac{l_{1}}{V(t)}\,y,t \right)
    \nonumber\\[6pt]
    \!&\sim&\! \left(\frac{l_{1}}{V(t)}\right)^{n}\!
    \frac{l_{1}}{|l_{1}|} \int_{-l_{1}\infty}^{l_{1}\infty}
    dy y^{n}e^{-y}\theta(y).
\label{Mn1}
\end{eqnarray}
Therefore, since $\int_{-l_{1}\infty}^{l_{1}\infty} dy y^{n}e^{-y} \theta(y) =
l_{1}n!/|l_{1}|$, the asymptotic formula (\ref{Mn1}) reduces to
\begin{equation}
    M_{n}(t) \sim \! \left( \frac{l_{1}}{V(t)} \right)^{\!n}n!.
\label{Mn2}
\end{equation}

An important feature of the moments of the particle position is that they are
slowly varying functions at infinity. This follows from the fact \cite{BGT}
that if some function [in our case $V(t)$] is slowly varying then so does any
power of this function. Due to this feature, in the long-time limit the
condition $M_{n}(t) /t^{\rho} \to 0$ holds for all $\rho>0$ \cite{BGT}. It
shows that the moments $M_{n}(t)$ grow to infinity (because $V(t)$ tends to
zero as $t \to \infty$) slower than any positive power of time. Hence, the CTRW
with superheavy-tailed distributions of waiting times can be viewed as a
generic model of superslow diffusion.

\subsection{Laws of superslow diffusion}
\label{Laws}

The character of diffusion is usually determined by the law of diffusion; that
is, the long-time behavior of the mean-square displacement or variance
\begin{equation}
    \sigma^{2}(t) = M_{2}(t) - M_{1}^{2}(t).
\label{var}
\end{equation}
Using the asymptotic formulas (\ref{M2n2}) and (\ref{Mn2}), from the definition
(\ref{var}) we find the diffusion laws
\begin{equation}
    \sigma^{2}(t) \sim \left\{\!\! \begin{array}{ll}
    l^{2}_{1}/V^{2}(t),
    & l_{1} \neq 0
    \\ [6pt]
    l_{2}/V(t),
    & l_{1} = 0
    \end{array}
    \right.
\label{var1}
\end{equation}
for both unbiased ($l_{1}=0$) and biased ($l_{1} \neq 0$) CTRWs with
superheavy-tailed distributions of waiting times. Since in this case the
variance $\sigma^{2}(t)$ is a slowly varying function, it grows to infinity
slower than any positive power of $t$, i.e., superslow diffusion occurs. As is
clear from Eq.~(\ref{var1}), the biased superslow diffusion is faster than the
unbiased one. We note that the same laws of superslow diffusion have also been
derived by using another approach \cite{DK2}, which does not involve explicitly
the long-time behavior of the probability density $P(x,t)$.

Although superslow diffusion has been observed in very different systems, all
previously known diffusion laws are given by a power function of the logarithm
of time: $\sigma^{2}(t) \propto \ln^{\nu} t$ ($\nu>0$). The best known example
of this type of diffusion is the Sinai diffusion \cite{Sin} for which $\nu=4$.
Other examples have been observed, e.g., in resistor networks \cite{HBSSB},
two-dimensional lattices \cite{BH,BO}, CTRWs \cite{HW1}, charged polymers
\cite{SSB}, aperiodic environments \cite{ITR}, iterated maps \cite{DrKl},
Langevin dynamics \cite{DH}, and fractional kinetics \cite{CKS}. In contrast,
the asymptotic formulas (\ref{var1}) define a large class of laws of superslow
diffusion, which includes $\sigma^{2}(t) \propto \ln^{\nu} t$ as a very
particular case. This class consists of the slowly varying functions that tend
to infinity as time evolves.

According to Eq.~(\ref{var1}), the diffusion laws are completely determined by
the asymptotic behavior of the exceedance probability $V(t)$. In particular,
for the examples considered in Sec.~\ref{Ex} the laws of superslow diffusion
can be obtained by the direct substitution of the asymptotic formulas
(\ref{V3as}), (\ref{V4as}) and (\ref{V5as}) into Eq.~(\ref{var1}). For the
first example we obtain
\begin{equation}
    \sigma^{2}(t) \sim \left\{\!\! \begin{array}{ll}
    \displaystyle \bigg(\frac{l_{1}\Gamma(1/q, b\ln^{q} \eta)}
    { b^{1/q -1}}\bigg)^{\!2} \frac{e^{2b\ln^{q}t}}{\ln^{2(1-q)}t},
    & l_{1} \neq 0
    \\ [12pt]
    \displaystyle \frac{l_{2}\Gamma(1/q, b\ln^{q} \eta)}
    { b^{1/q -1}}\, \frac{e^{b\ln^{q}t}}{\ln^{1-q}t},
    & l_{1} = 0.
    \end{array}
    \right.
\label{var2}
\end{equation}
The second example leads to the following law of superslow diffusion:
\begin{equation}
    \sigma^{2}(t) \sim \left\{\!\! \begin{array}{ll}
    \displaystyle \bigg( \frac{l_{1}}{\ln^{r-1}\eta}
    \bigg)^{\!2}\ln^{2(r-1)}t,
    & l_{1} \neq 0
    \\ [12pt]
    \displaystyle \frac{l_{2}}{\ln^{r-1}\eta}\, \ln^{r-1}t,
    & l_{1} = 0,
    \end{array}
    \right.
\label{var3}
\end{equation}
which for $l_{1}=0$ was first derived in Ref.~\cite{HW1}. In this example, that
reproduces the diffusion laws with $\sigma^{2}(t) \propto \ln^{\nu} t$,
diffusion occurs slower than in the previous one. Finally, the third example
yields
\begin{equation}
    \sigma^{2}(t) \sim \left\{\!\! \begin{array}{ll}
    \displaystyle \bigg( \frac{l_{1}}{(\ln\ln \eta)^{r-1}}
    \bigg)^{\!2}(\ln\ln t)^{2(r-1)},
    & l_{1} \neq 0
    \\ [12pt]
    \displaystyle \frac{l_{2}}{(\ln\ln \eta)^{r-1}}\,
    (\ln\ln t)^{r-1},
    & l_{1} = 0,
    \end{array}
    \right.
\label{var4}
\end{equation}
i.e., the variance grows slower than that in Eq.~(\ref{var3}). We recall that
these examples are illustrative because infinitely many different laws of
superslow diffusion exist.

\subsubsection{Specific application}
\label{Appl}

Now, as an application of our results, we will show that particles moving under
a constant force $f(>0)$ through a randomly layered medium can exhibit the
biased superslow diffusion. According to \cite{DK3}, the influence of this
medium on the particle dynamics can be described by the piecewise constant
random force $g(x)= g^{(n)}$, where $x \in [nl,(n+1)l)$, $n = 0,1,\ldots$, $l$
is the layer thickness, and $g^{(n)}$ are independent random variables
distributed with the probability density $u(g)$ in the interval $[-g_{0},
g_{0}]$. In the overdamped regime the particle position $x_{t}$ ($x_{0}=0$) is
governed by the equation of motion $\nu \dot{x}_{t} = f + g(x_{t})$ with $\nu$
being the damping coefficient. At $f\geq g_{0}$ the long-time solution of this
equation can be approximated by the walker position $X(t)$ if the jump density
is given by $w(x) = \delta(x-l)$ and the waiting-time density $p(\tau)$ is
associated with the probability density of the random time $\tau^{(n)}$ that a
particle spends moving from the point $nl$ to the point $(n+1)l$
\cite{DK3,DDK}. Since $\tau^{(n)} = \nu l/(f + g^{(n)})$, the probability
density $u(g)$ which characterizes the medium disorder is expressed through
$p(\tau)$ as follows:
\begin{equation}
    u(g) = \frac{\nu l}{(f + g)^{2}}\, p \bigg(\frac{\nu l}
    {f + g}\bigg).
\label{u(g)}
\end{equation}
It should be noted that the waiting-time density $p(\tau)$ is defined in the
interval $[\tau_{\mathrm{min}}, \tau_{\mathrm{max}}]$, where $\tau_{
\mathrm{min}} = \nu l/(f + g_{0})$ and $\tau_{\mathrm{max}} = \nu l/(f -
g_{0})$.

It was shown in Ref.~\cite{DDK} that the biased diffusion of particles can be
anomalous only if $f=g_{0}$. In this case the character of anomalous diffusion
is determined by the behavior of $p(\tau)$ as $\tau \to \infty$ (we recall that
$\tau_{\mathrm{max}} = \infty$ at $f=g_{0}$) or, as it follows from
Eq.~(\ref{u(g)}), by the behavior of $u(g)$ as $g \to -g_{0} + 0$. Thus, taking
into account the asymptotic formula (\ref{p as}), one can see that if
\begin{equation}
    u(g) \sim \frac{1}{g_{0} + g}\, h \bigg(\frac{1}
    {g_{0} + g}\bigg)
\label{u(g) as}
\end{equation}
($g \to -g_{0} + 0$) then the biased diffusion of particles moving under a
force $f=g_{0}$ in a randomly layered medium is superslow with $\sigma^{2}(t)
\sim l^{2}/V^{2}(t)$ as $t \to \infty$ [since $w(x) = \delta(x-l)$, we have
$l_{1} = l$]. In particular, Eq.~(\ref{u(g) as}) for $p(\tau)$ from
Eq.~(\ref{p2}) yields
\begin{equation}
    u(g) \sim \frac{(r-1)\ln^{r-1}\eta}{(g_{0} + g)
    |\ln (g_{0} + g)|^{r}},
\label{u(g) as2}
\end{equation}
and so the law of superslow diffusion in such a medium is given by
Eq.~(\ref{var3}) with $l_{1} = l \neq 0$.

\section{CONCLUSIONS}
\label{Concl}

We have studied the long-time behavior of the decoupled CTRW with a
superheavy-tailed distribution of waiting times. We assume that the
distribution function of waiting times belongs to a class of slowly varying
functions and the jump distribution has a finite second moment. At these
conditions, the limiting probability density of the properly scaled particle
position has been derived for both unbiased and biased CTRWs. In the former
case the limiting density is given by the symmetric two-sided exponential
density, while in the latter case it takes the form of the asymmetric one-sided
exponential density. We have used these limiting densities to calculate the
moments of the particle position at long times. It has been shown that all the
moments are expressed through the exceedance probability and grow slower than
any positive power of time.

Due to the last property, the decoupled CTRW with a superheavy-tailed
distribution of waiting times can be viewed as a generic model of superslow
diffusion. We have derived the most general form of the laws of superslow
diffusion and have shown that, in the cases of unbiased and biased diffusion,
the variance of the particle position is inversely proportional to the first
and second powers of the exceedance probability, respectively. The laws of
superslow diffusion that correspond to the illustrative examples of
superheavy-tailed distributions of waiting times have also been determined.
Finally, we have applied the obtained results to show that the biased diffusion
of particles moving under a constant force in a randomly layered medium can be
superslow.

\end{document}